\documentclass[seceq]{ptptex}

\usepackage[usenames]{color}
\usepackage[dvips]{epsfig}
\usepackage{amssymb}

\usepackage{wrapft}

\def\lsim{\raise0.3ex\hbox{$<$\kern-0.75em\raise-1.1ex\hbox{$\sim$}}}
\def\gsim{\raise0.3ex\hbox{$>$\kern-0.75em\raise-1.1ex\hbox{$\sim$}}}
\title{%        %You can use \\ for explicit line-break.
O(N) universality and the chiral phase transition in QCD%
}

\author{%       %Use \scshape for the family name.
Frithjof \textsc{Karsch}%
}

\inst{%     %Affiliation, neglected when [addenda] or [errata].
Physics Department, Brookhaven National Laboratory,Upton, NY 11973, USA\\
and\\
Fakult\"at f\"ur Physik, Universit\"at Bielefeld, D-33615 Bielefeld, Germany
}

\abst{
We discuss universal scaling properties of (2+1)-flavor QCD  in the vicinity 
of the chiral phase transition at
vanishing as well as non-vanishing light quark chemical potential ($\mu_l$). 
We provide evidence for $O(N)$ scaling of the chiral order parameter in
(2+1)-flavor QCD and show that the scaling analysis of its derivative
with respect to the light quark chemical potential provides a unique
approach to the determination of the curvature of the chiral phase transition
line in the vicinity of $\mu_l/T=0$.
}

\begin{document}

\maketitle

\section{Introduction}

It is well understood that the phase structure of QCD at non-zero 
temperature crucially depends on the quark mass values 
\cite{Pisarski,Columbia}. While in the
case of three degenerate quark masses (3-flavor QCD) a first order
transition occurs for sufficiently small values of the quark mass,
it is expected that the transition in 2-flavor QCD is continuous
and belongs to the universality class of $O(4)$ symmetric, 3-dimensional 
spin models. Several studies of the QCD phase diagram as function of two 
degenerate light ($m_u\ ,m_d$) and a strange ($m_s$) quark mass suggest 
that the region around the 3-flavor chiral limit, where the QCD transition 
becomes first order, is indeed small and does not include the point of 
physical light and strange quark masses. Our 
\begin{wrapfigure}{l}{\halftext}
\epsfig{file=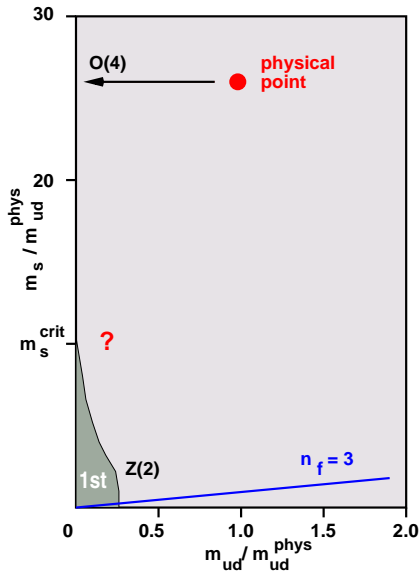,width=5.5cm}
\caption{The phase diagram of 3-flavor QCD.
}
\label{fig:massplane}
\end{wrapfigure}
\noindent
current understanding of
the location of the region of $1^{st}$ order transitions that is separated
by a line of $2^{nd}$ order transitions from the crossover region and its
positioning relative to the physical point is
presented in the Columbia plot shown in Fig.~\ref{fig:massplane}.

While numerical calculations in 3-flavor QCD gave evidence for the
existence of a first order transition,
many of the details of the transition in 2- or (2+1)-flavor QCD
with light up and down quarks are still poorly constrained through lattice
calculations. Earlier attempts to verify $O(4)$ scaling in 
numerical studies of 2-flavor QCD using standard staggered fermions were not 
very successful in establishing the expected universal scaling properties \cite{earlier}.

\noindent\hspace*{-7.0cm}
We present here a new analysis of scaling properties using ${\cal O}(a^2)$
improved staggered fermion and gauge actions \cite{ourON}. 
We perform this analysis for
(2+1)-flavor QCD. Implicitly we thus also analyze whether the region of first order
chiral transitions ends in a tri-critical point for a strange quark mass value
below its physical value or above. We will use the scaling analysis of the chiral
order parameter to determine the curvature of the chiral phase transition line
at non-vanishing values of the light quark chemical potential $\mu_l$, {\it i.e.}
the first (Taylor) expansion coefficient of the transition line in powers 
of $(\mu_l/T)^2$ around $\mu_l/T=0$.

\section{O(N) scaling of the chiral condensate}

In the vicinity of a critical point regular contributions to the logarithm of the
partition function become negligible.
The universal critical behavior of the order parameter $M$ of, e.g. 3-dimensional
$O(N)$ spin models, is then controlled by a scaling function $f_G$ that arises from
the singular part of the logarithm of the partition function,
\begin{equation}
M(t,h) \;=\; h^{1/\delta} f_G(z) \; ,
\label{order}
\end{equation}
with  $z=t/h^{1/\beta\delta}$ and scaling variables $t$ and $h$ that are related to
the temperature, $T$, and the symmetry breaking (magnetic) field, $H$,
\begin{equation}
t = \frac{1}{t_0}\frac{T-T_c}{T_c} \quad , \quad h= \frac{H}{h_0} \; .
\label{reduced}
\end{equation}
Here $\beta$ and $\delta$ are critical exponents, unique for the universality 
class of the second order phase transition which the system undergoes in the
limit $(t,h)\rightarrow (0,0)$. The form of the scaling function $f_G(z)$ is well
known from numerical simulations of 3-dimensional $O(N)$ symmetric spin models
\cite{Engels}.

In QCD symmetry breaking arises due to a non-vanishing light quark mass,
$H\equiv m_l/m_s$, and the corresponding order parameter is the chiral condensate, 
which we write as
\begin{equation}
M\equiv M_b = N_\tau^4 m_s \langle \bar{\psi}\psi \rangle_l \; ,
\label{Mb}
\end{equation}
where $m_s$ denotes the strange quark mass in lattice units and $N_\tau$ is the
temporal extent of the 4-dimensional lattice, $N_\sigma^3\times N_\tau$. 
One may 
improve the operator $M_b$ by subtracting a fraction of the strange quark condensate
\cite{ourON}. This eliminates additive linear divergent terms proportional to 
the light quark masses. Such a subtraction is, in fact, mandatory, if one wants
to take the continuum limit at finite quark mass before taking the chiral limit.
This ordering of limits, indeed is needed in order to recover the correct 
$O(4)$ scaling behavior from calculations performed with staggered fermions, 
which only preserve a global $O(2)$ symmetry for any non-zero value of the lattice 
spacing.
We will be less ambitious here and discuss the chiral limit on lattices with
fixed temporal extent, $N_\tau=4$. 
In this case, we can only expect to find 
$O(2)$ rather than $O(4)$ scaling behavior. However, the scaling functions, $f_G(z)$, 
are very similar for both universality classes and it thus will be difficult to
distinguish $O(2)$ and $O(4)$ scaling through an analysis of the order parameter 
alone. Moreover, given the large scaling violations
observed in earlier studies with staggered fermions\cite{earlier}, already
the observation of scaling in terms of a {\it generic O(N)} scaling function
at non-zero values of the lattice cut-off is a major step forward. 

\begin{figure}[t]
\begin{center}
\epsfig{file=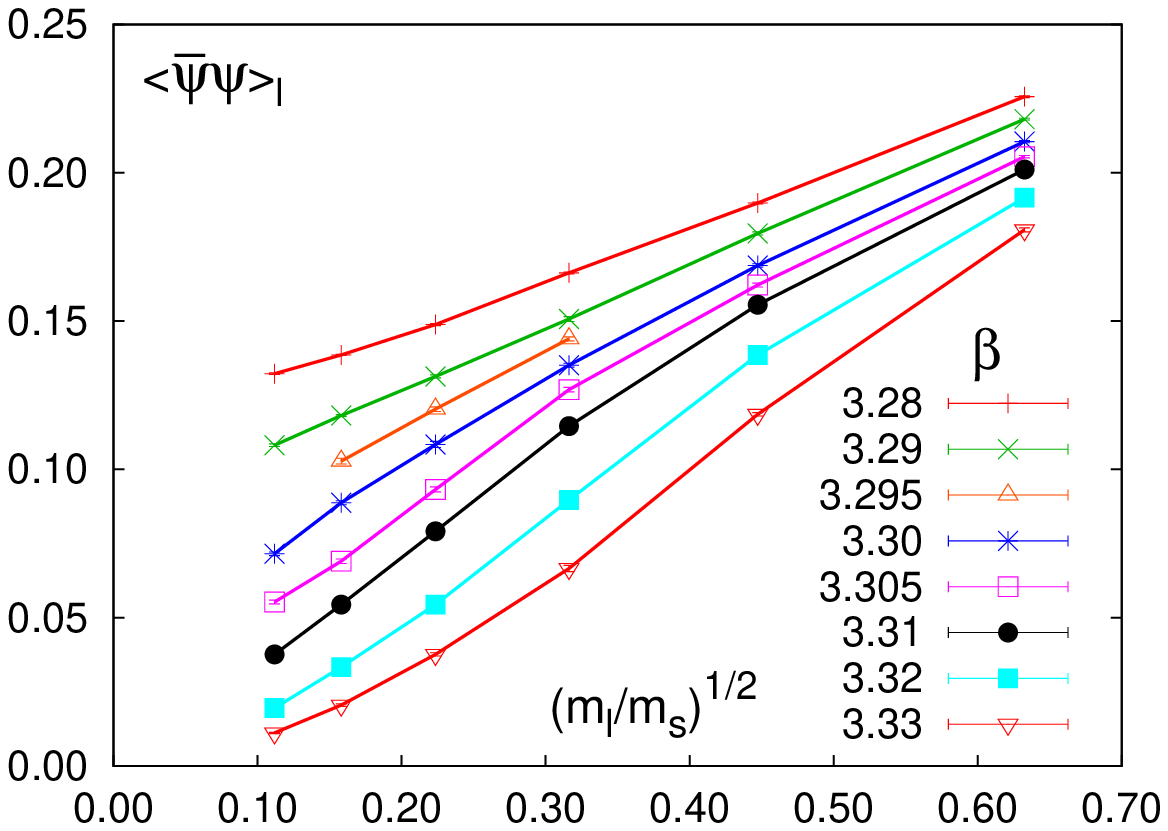,width=7.4cm}\hspace*{-0.4cm}\epsfig{file=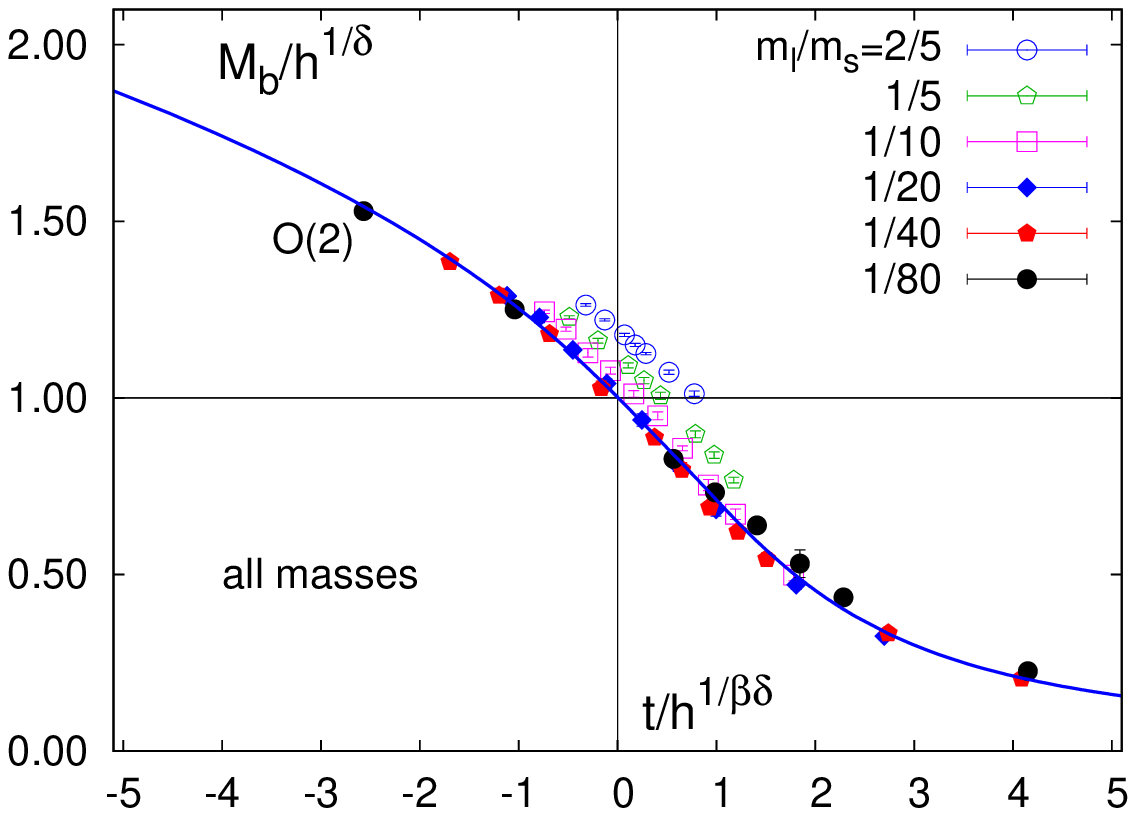,width=7.4cm}
\end{center}
\caption{\label{fig:pbp}The light quark chiral condensate in lattice units versus
the ratio of the square root of the light and strange quark masses (left) and its 
scaling form (right). The right hand figure has been obtained using fits for
$m_l/m_s \le 1/20$ and $|T-T_c|/T_c\le 1.03$ only.
}
\end{figure}

We show in Fig.~\ref{fig:pbp}(left) results from a calculation of chiral
condensates in (2+1)-flavor QCD. The bare strange quark mass ($m_s$) has been 
chosen such that the physical value of the strange pseudo-scalar mass,
$m_{s\bar{s}}\equiv \sqrt{2m_K^2 - m_\pi^2}$, is reproduced.
The light quark mass  
has been varied in a range $1/80\le m_l/m_s\le 2/5$, which 
for the light pseudo-scalar Goldstone meson corresponds to a regime 
$75 {\rm MeV} \lsim m_{ps} \lsim 420{\rm MeV}$. The lattice size has been varied
from $16^3\times 4$ for the heavier quark masses to $32^3\times 4$ for the 
lightest quark masses \cite{ourON}. This insures that finite volume effects remain 
small in the entire light quark mass regime, {\it i.e.} in units of the spatial
extent $N_\sigma$  we always have $m_{ps}N_\sigma \ge 3$.

From Fig.~\ref{fig:pbp}(left) it is obvious that the chiral condensate
scales with the square root
of the quark mass in the low temperature, chiral symmetry broken phase, 
\begin{equation}
\langle \bar{\psi}\psi \rangle = a(T)+b(T)\sqrt{\frac{m_l}{m_s}} + 
{\cal O}\left ( \frac{m_l}{m_s} \right) \ .
\label{psibarpsi}
\end{equation}
This is characteristic for Goldstone-modes in three dimensional $O(N)$ 
symmetric spin models. In fact, this also is the dominant term characterizing
the scaling function $f_G(z)$ in the symmetry broken phase, {\it i.e.} for
$z<0$ \cite{ourON}.

Results for the chiral condensate may be put on the universal scaling curve by using 
the reduced temperature and rescaled symmetry breaking field introduced in 
Eq.~\ref{reduced}. Of course, this is expected to be possible only 
close to criticality where contributions from regular
terms and corrections to scaling are small.
The scaling analysis shown in Fig.~\ref{fig:pbp}(right) has therefore only been
performed for the three lightest quark mass values, $m_l/m_s \le 1/20$ and for 
temperatures close to $T_c$. From this
one determines the three free parameters, $t_0,\ h_0$ and $T_c$. As expected results 
for heavier quarks, also shown in Fig.~\ref{fig:pbp}(right), show deviations
from the universal scaling behavior. Contributions from corrections to scaling become 
significant for $m_l/m_s\gsim 1/5$, {\it i.e.} $m_{ps}\gsim 300$~MeV. This is in
contrast to calculations with Wilson fermions, where indications for $O(4)$ 
scaling have been reported for even large values of the pseudo-scalar meson mass
\cite{Wilson}.
 
The scaling analysis of the order parameter provides two non-universal 
parameters that are unique for QCD, the chiral phase transition temperature, $T_c$,
and the scale parameter $z_0 = h_0^{1/\beta\delta}/t_0$.  In the
continuum limit both quantities are functions of the strange quark mass only.
Of course, $T_c$ as well as $z_0$ are cut-off dependent and a proper continuum
extrapolation is needed to extract their values in the continuum limit.
From our analysis on lattices with temporal extent $N_\tau= 4$ we find
$z_0 \simeq 7.5$. A preliminary analysis on lattices with temporal extent $N_\tau =8$
suggests that this value drops by almost a factor 2 \cite{Unger}. A more detailed
analysis of the approach to the continuum limit thus is needed. We stress,
however, that $z_0$ is a physical parameter of QCD. It 
gives the slope of the quark mass dependence of the 
pseudo-critical temperature, which can be  determined from the location of
a peak in the chiral susceptibility,
\begin{eqnarray}
\chi_M(t,h) &=& \frac{\partial M}{\partial H} = \frac{1}{h_0}
h^{1/\delta -1} f_\chi(z) \; , \label{chiM}\\
f_\chi(z) &=& \frac{1}{\delta}\left( f_G(z) -\frac{z}{\beta} f'_G (z)\right) \; .
\label{fchi}
\end{eqnarray}
The scaling functions $f'_G(z)={\rm d}f_G/{\rm d}z$ and $f_\chi(z)$ are shown 
in the right hand part of Fig.~\ref{fig:c2pbp}. 
The scaling function $f_\chi(z)$ has a maximum at $z_p$. 
The dependence of the pseudo-critical temperature, $T_p$,
on the quark mass is given by the condition that 
$z=t/h^{1/\beta\delta} =z_p$, {\it i.e.}
\begin{equation}
\frac{T_{p}(H) - T_c}{T_c} = \frac{z_p}{z_0}
H^{1/\beta\delta} \ .
\label{pseudo}
\end{equation}
For the 3-d $O(2)$ universality class the
peak in the chiral susceptibility is located at $z_p \simeq 1.56$. 
Using this and expressing the symmetry breaking field $H$ in terms
of pion and kaon masses rather than quark masses,
$H=m_l/m_s \simeq 0.52 \, (m_{ps}/m_K)^2$, we find
\begin{equation}
\frac{T_{p}(m_{ps})}{T_c} = 1+ \frac{1.06}{z_0} 
\left( \frac{m_{ps}}{m_K} \right)^{2/\beta\delta} \ .
\label{pseudo2}
\end{equation}
This allows to estimate the phase transition temperature in the chiral limit. With 
$m_\pi/m_K \simeq 0.27 $ one finds with our current estimates for $z_0$ that the 
transition temperture in the chiral limit is about $7\%$ smaller than the crossover 
temperature at
physical values of the quark masses. Of course, this still needs to be analyzed closer 
to the continuum limit.
 
\section{Curvature of the critical line in the \boldmath $T$-$\mu_B$ phase diagram}

For non-vanishing light quark chemical potential, $\mu_l$, the second order chiral
phase transition persists to exist in the $T$-$\mu_l$ plane. For
small values of the chemical potential, $\mu_l/T\gsim 0$, the curvature of the phase
transition line can be determined by making use of the scaling analysis of the
order parameter. In this case the reduced temperature variable, $t$,
also depends on the chemical potential as it couples to the quark number operator, which
does not break chiral symmetry. To leading order it contributes quadratically,
\begin{equation}
t=\frac{1}{t_0}\left( \frac{T-T_c}{T_c} +\kappa_\mu \left( \frac{\mu_l}{T}\right)^2\right) \ .
\label{reducedmu}
\end{equation}

\begin{wrapfigure}{l}{\halftext}
\begin{center}
\vspace*{-0.9cm}
\epsfig{file=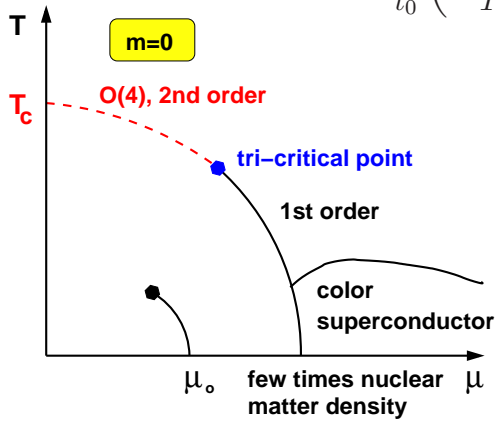,width=6.5cm}
\end{center}
\caption{\label{fig:curvature} The curvature of the transition curve in the $T$-$\mu_B$
plane. 
}
\end{wrapfigure}

\begin{figure}[t]
\begin{center}
\epsfig{file=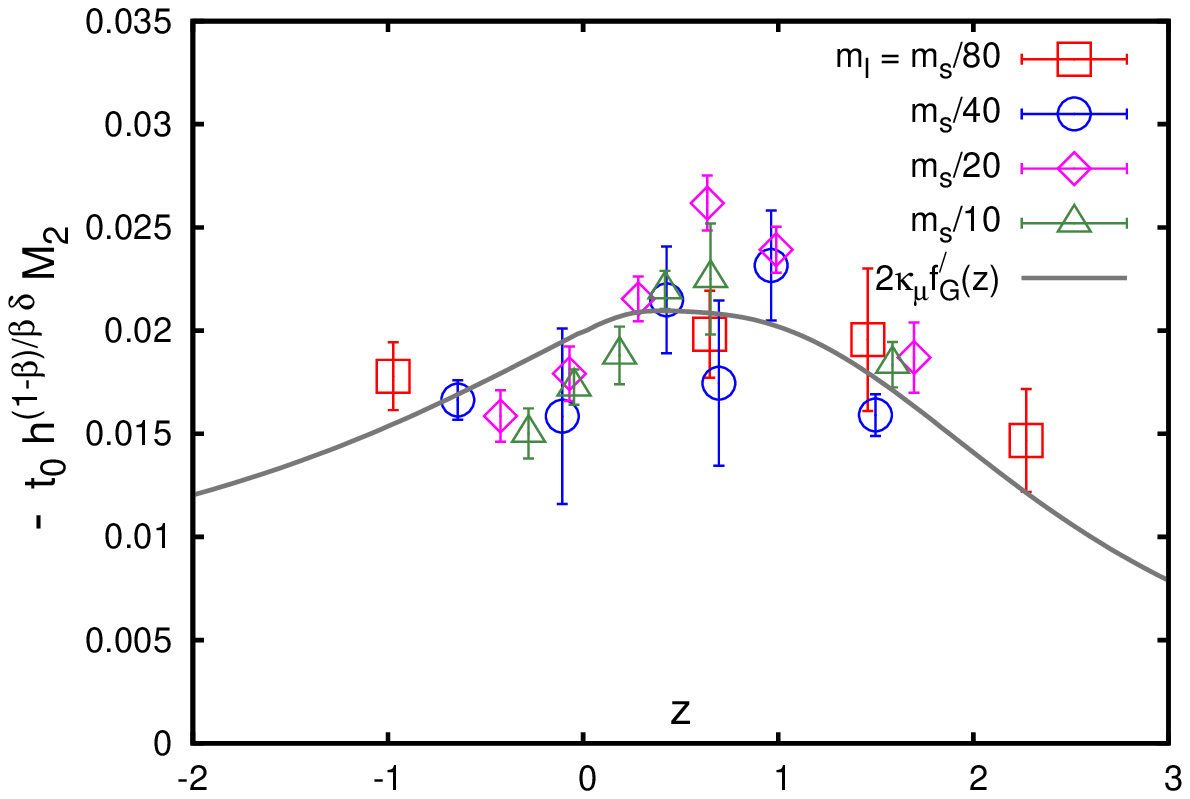,width=7.1cm}
\hspace*{-0.4cm}\epsfig{file=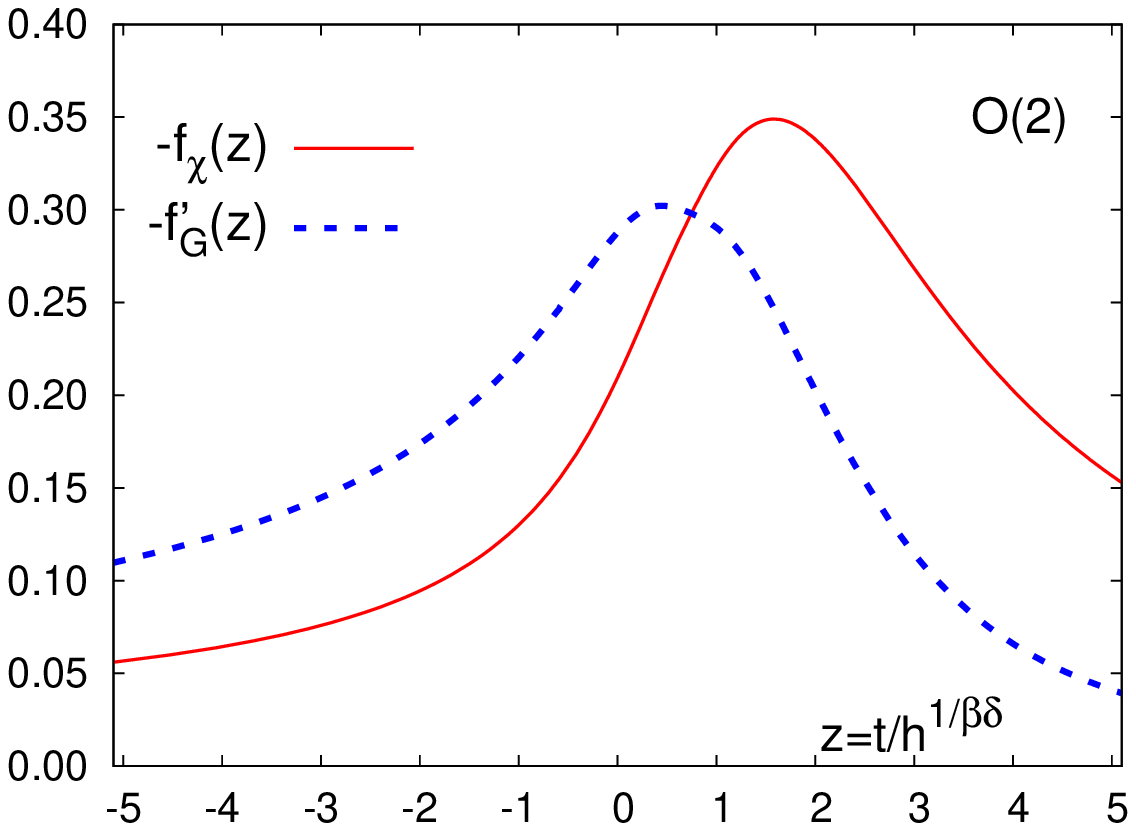,width=6.9cm}
\end{center}
\caption{\label{fig:c2pbp} The second derivative of the order parameter with
respect to the light quark chemical potential (left). The  $O(2)$ scaling
curve $f'_G(z)$, shown together with $f_\chi(z)$ in the right hand part of
the figure, has been rescaled to match the data.
}
\end{figure}

\noindent
The condition for criticality, $t=0$, fixes the shape of the transition line
for small values of $\mu_l/T$.
In the scaling regime the order parameter $M$ depends on temperature and quark mass
only through the scaling
variable $z$.
The derivative of $M$ with respect to $T$ thus 
is, up to a constant, identical to the second derivative of $M$ with respect to 
the chemical potential $\mu_l$. This derivative too is related to the scaling 
function $f'_G(z)$ which we have introduced in Eq.~\ref{fchi} and which is
shown in Fig.~\ref{fig:c2pbp}(right),

\begin{eqnarray}
M_2 &\equiv& \left. 
\frac{\partial^2 M}{\partial (\mu_l/T)^2}\right|_{\mu_l=0} =
\frac{2 \kappa_\mu}{t_0} h^{(\beta-1)/\beta\delta} f'_G(z) \ ,
\label{scalingmu} 
\end{eqnarray} 
We note that $M_2$ has properties similar to the chiral susceptibility. It
diverges in the chiral limit at $z=0$ and the peak position in $f'_G(z)$
can be used to define a pseudo-critical temperature at non-zero values of the 
light quark mass. 

Once the scale parameters $t_0,\ h_0, \ T_c$, needed to project the chiral order 
parameter onto the $O(N)$ scaling curve, are known, we can use this information to 
determine the curvature of the critical line, $\kappa_\mu$, by calculating 
$M_2$ and by matching the scaling curve to the data using an appropriate
scaling factor $\kappa_\mu$. 
Such a scaling analysis is shown in Fig.~\ref{fig:c2pbp}(left).
From this we find for the curvature of the phase transition line in the chiral limit
the preliminary result $\kappa_\mu \simeq 0.035$ \cite{ourkappa},
which is in good agreement with earlier determinations of the curvature 
of the pseudo-critical line at non-zero values of the quark mass performed with 
different numbers of flavors \cite{Philipsen}. 
We note that the current analysis suggests that the curvature of the 
transition line, expressed in terms 
of\footnote{Some caution is needed when translating $\mu_l$ to $\mu_B$ and 
comparing lattice results with experimental findings. Lattice calculations
have been performed for vanishing strange quark chemical potential (in some cases
even for 2-flavor QCD), while the
freeze-out conditions in a heavy ion collision refer to a system with 
$\mu_s/\mu_B \simeq 0.16$ \cite{cleymans}.}
$\mu_B \sim 3 \mu_l$, is smaller than the experimentally
determined freeze-out curve, which is well parametrized by \cite{cleymans},
\begin{equation}
\frac{T}{T_{\rm freeze-out}} \simeq 1 - 0.023 \left( \frac{\mu_B}{T} \right)^2 +
{\cal O}\left( \frac{\mu_B}{T} \right)^4 \ .
\label{freezeout}
\end{equation}

\section{Conclusions}
\vspace*{-0.1cm}
We have discussed universal properties of the chiral condensate of (2+1)-flavor
QCD in the limit of vanishing light quark masses. It agrees well with expected
$O(N)$ scaling predictions. We showed how this can
be used to calculate the curvature of the second order chiral phase transition
line in the vicinity of $\mu_B/T=0$. 

At present this analysis is limited to rather coarse lattices. Calculations closer
to the continuum limit are needed to get control over the scale parameter $z_0$
and the phase transition temperature in the chiral limit.

\section*{Acknowledgements}
\vspace*{-0.3cm}
We thank the organizers of the workshop 'New Frontiers in QCD 2010' at 
the Yukawa Institute for Theoretical Physics, Kyoto, for a very stimulating 
workshop program and for support. 
This work has been supported in part by contract DE-AC02-98CH10886 with the
U.S. Department of Energy.

\end{document}